%% file: main.tex
\documentclass[sigconf,nonacm]{acmart}

\setcopyright{none}
\AtBeginDocument{%
  }

\usepackage{hyperref}
\usepackage{enumitem}
\usepackage[ruled,linesnumbered]{algorithm2e}
\usepackage{multicol}
\usepackage{multirow}
\usepackage{makecell}
\usepackage{tabularray}
\usepackage{tabularx}
\usepackage{kotex}
\usepackage{pifont}
\usepackage{xcolor}
\usepackage{hhline}
\usepackage{subcaption}
\usepackage{booktabs}
\usepackage{verbatim}
\usepackage[utf8]{inputenc}
\usepackage[T1]{fontenc}
\usepackage{subcaption}
\usepackage{caption}
\usepackage[
  separate-uncertainty = true,
  multi-part-units = repeat
]{siunitx}
\usepackage{algorithmic}
\usepackage{microtype}
\usepackage{tabularx}

\SetKwFor{ForAll}{for all}{do}{}
\SetKwFor{For}{for}{do}{}
\SetKwFor{ForParallel}{for}{do in parallel}{}
\SetKwFor{While}{while}{do}{}
\SetKwIF{If}{ElseIf}{Else}{If}{then}{ElseIf}{Else}{}
\SetKwComment{Comment}{/* }{ */}
\SetKwFor{ForEach}{for each}{do}{}

\begin{document}

\title{AkasicMEM: Governed Enterprise Memory for Agents}
\renewcommand{\shorttitle}{GraphAI}

\author{Jeongmin Bae, \ Yongjae Kim, \ Kyoung Hur, \ Donghyoung Han, \ Min-Soo Kim}
\affiliation{
  \institution{\textbf{GraphAI}}
  \country{Seoul, Republic of Korea}
}

\renewcommand{\shortauthors}{GraphAI}

\begin{abstract}
Agent memory enables enterprise agents to retain knowledge acquired during work and reuse it across tasks and agents, turning execution experience into persistent organizational knowledge.
Realizing this potential requires both source--memory integration, through which enterprise sources and accumulated memory can be utilized together, and memory governance, through which shared memory remains subject to organizational policies throughout its lifecycle.
These requirements interact when information from enterprise sources persists in memory.
As this information is repeatedly derived and reused under changing principals and policies, source restrictions may be bypassed, resulting in information leakage.
Preventing such leakage requires authorization continuity, under which source restrictions remain effective throughout source-to-memory and memory-to-memory derivation and reuse.
Existing approaches address these concerns individually, but do not treat source--memory integration, memory governance, and authorization continuity as combined design targets across the memory lifecycle.
We define \textit{Governed Enterprise Memory} as agent memory designed around this combined scope and present \textit{AkasicMEM} as its realization.
AkasicMEM realizes authorization continuity through transitive lineage, policy composition during memory formation, and policy re-evaluation during retrieval.
It is built on GraphAI's AkasicDB, a unified vector--graph--relational database whose storage and execution substrate enables the underlying operations of these mechanisms to be jointly optimized and executed.
\end{abstract}

\keywords{Agent memory, Enterprise data, Governed enterprise memory}

\maketitle

\input{1_introduction}

\input{2_background}
\input{3_section}
\input{4_AkasicMEM}
\input{5_discussion}
\input{6_conclusion}

\bibliographystyle{ACM-Reference-Format}
\bibliography{References}

\end{document}

%% file: 1_introduction.tex
\section{Introduction}
\label{sec:introduction}

Agent memory enables AI agents to retain knowledge and experience acquired through interactions with users and task execution, and to reuse them in subsequent tasks~\cite{agent_memory}.
Existing agent memory systems, including Mem0~\cite{mem0} and Letta (formerly MemGPT)~\cite{memgpt}, have primarily been developed for personalized assistants.
They retain facts and preferences about users and preserve conversational context across sessions.

In enterprise settings, interest has grown in agent memory that goes beyond personalized context and lets knowledge acquired during actual work be reused across tasks and agents.
Organizations can package established methods and procedures as agent skills~\cite{agent_skill}, but skills reuse knowledge that has been specified explicitly and cannot capture what is learned during execution.
Agent memory complements them by accumulating newly discovered facts, decision context, and execution outcomes for subsequent use.
In practice, SAP researchers report a deployment in which organization-specific knowledge gained during software development is accumulated as memory and provided to subsequent coding agents~\cite{sap}.
G-Memory similarly enables the collaboration processes of multiple agents and the lessons learned from them to be reused in later tasks~\cite{g-memory}.
These studies demonstrate the potential of agent memory to turn experience from task executions into persistent organizational knowledge.

Realizing these benefits in enterprise settings, however, requires that \textbf{agent memory support both \textit{source--memory integration} and \textit{memory governance}.}
Source--memory integration enables information maintained in existing business systems to be used together with memory formed through prior agent executions.
We refer to customer and contract records, internal documents, and other information maintained in enterprise systems as \textit{enterprise sources}.
This integration brings enterprise source representations and agent memory into a common data management and retrieval architecture, while preserving the distinction between source data and derived memory.
Because enterprise sources can describe current business states and rules, whereas memory preserves knowledge obtained from previous tasks, task-relevant context must be constructed from both.
Retrieval over enterprise sources, commonly implemented through RAG~\cite{rag}, must therefore be coordinated with memory formation and retrieval.

Memory governance concerns applying access policies, lifecycle management, and auditability to memory shared across users and agents.
In an enterprise, sharing memory does not mean making all information universally accessible.
It means that authorized users and agents can reuse knowledge and experience within their permitted scope.
A \textit{memory scope} specifies the user, agent, thread, project, team, or tenant boundary within which a memory may be retrieved and reused.
Because these boundaries can change with organizational roles and business relationships, governance must address not only access to memory but also its creation and modification, retention and deletion, and disclosure~\cite{collaborative_memory, oracle, caura_ai}.

These two requirements interact when enterprise sources contribute to memory formation.
Memory derived from enterprise sources is reused and in turn contributes to new memory, and over that time both the principals requesting it and the policies of its sources may change.
A memory scope determines the intended boundary of reuse, but membership in that scope does not establish that a principal is authorized to access the information inherited from the contributing sources.
If a memory remains retrievable on the basis of its scope alone after access to a contributing source has been withdrawn, \textbf{the source restriction is bypassed through memory, resulting in information leaks}~\cite{collaborative_memory,map_graph}.
Source-derived restrictions must therefore remain effective as memory is repeatedly derived and reused under changing policies and principals, a requirement we call \textit{authorization continuity}.
Propagating source restrictions to derived information has long been studied in information flow control~\cite{difc}.
Agent memory extends the problem to memory that is continuously formed and reused, and whose derivation relationships and applied policies evolve over time.


Prior work has addressed each of these aspects, and some address more than one
(Section~\ref{sec:existing_memory_systems}), but to the best of our knowledge, none
treats source--memory integration, memory governance, and authorization continuity
as combined core design targets across the memory lifecycle.
We define a system designed for this combined scope as \textit{Governed Enterprise Memory} and introduce \textbf{AkasicMEM} as its realization.
AkasicMEM incorporates transitive lineage, policy composition during memory formation, and retrieval-time policy re-evaluation into the memory lifecycle to maintain authorization continuity, and records each declassification and retrieval decision for audit.

The technical foundation of AkasicMEM is \textit{AkasicDB}, GraphAI's unified vector, graph, and relational database~\cite{akasicdb}.
Governed Enterprise Memory requires vector search for semantic retrieval, graph processing for lineage and organizational authorization relationships, and relational processing for structured enterprise sources and policies.
AkasicDB supports all three natively on a single execution substrate, which extends a graph–relational query processor introduced in prior work~\cite{chimera} that unifies traversal and join processing, adding vector search so that retrieval, lineage traversal, and policy evaluation are processed together in a single plan.
This paper defines Governed Enterprise Memory and the requirements for preserving source authorization across memory derivation and reuse.
We present AkasicMEM, which records lineage and evaluates authorization policies during memory formation and retrieval on AkasicDB.
Finally, we discuss broader issues concerning the governance, usefulness, and evaluation of agent memory in enterprise settings.

%% file: 2_background.tex
\section{Background and Related Work}

\subsection{Agent Memory and Its Lifecycle}
\label{sec:agent_memory_lifecycle}

Agent memory is the information state that an AI agent maintains across reasoning steps, interactions, and tasks, and draws on in later decisions~\cite{agent_memory}.
Following the cognitive-architecture view of language agents~\cite{coala}, we distinguish \emph{working memory}, which holds the state of the current task, from \emph{long-term memory}, which persists beyond it.
Long-term memory comprises \emph{semantic memory} (facts about users, the organization, and the environment), \emph{episodic memory} (past interactions and task experiences), and \emph{procedural memory} (how tasks are performed).
We call an explicitly represented unit of such information that a system stores and manages a \emph{memory object}.
An \emph{agent memory system} is the software layer that manages memory objects throughout their lifecycle.

Agent memory is not a static archive but has a lifecycle in which memory is formed, changes over time, and can eventually be removed.
Recent surveys have termed this lifecycle in different ways, as dynamics of formation, evolution, and retrieval~\cite{agent_memory}, as extraction, storage, retrieval, and evolution~\cite{graph_memory_survey}, or as explicit memory actions such as store, update, evict, and delete~\cite{anatomy_agentic_memory}.
We adopt four dynamics, formation, retrieval, evolution, and forgetting, and unlike prior formulations we treat forgetting as a dynamic in its own right rather than as part of evolution.
This distinction is because, in an enterprise, a memory may be removed because a retention period ends, a contract is terminated, or a source is revoked.
Each of these is an operation with its own trigger and obligations.
Formation turns information worth keeping, such as a fact learned from a tool result, into a stored memory object.
Retrieval brings memory relevant to the current task into the agent's context, and must also check that the memory is still valid, because memory outlives the session that formed it.
A memory may have gone stale or been contradicted by newer information, and memory written through ordinary interaction, including content injected without any elevated privilege, is retrieved in later sessions as trusted context~\cite{agentpoison, minja}.
Evolution revises existing memory as new information arrives, and forgetting excludes from use or removes altogether memory that is no longer valid or need not be retained.
Around these dynamics runs the agent-level loop of observing, acting on retrieved memory, and observing the outcome, which feeds the next formation.

\begin{table*}[!t]
\centering
\caption{Comparison of existing agent memory systems with respect to the requirements of enterprise agent memory, grouped by the boundary across which memory is reused.}
\label{tab:existing_memory_systems}
\vspace{-8pt}
\footnotesize
\setlength{\tabcolsep}{2pt}
\renewcommand{\arraystretch}{1.1}
\begin{tabularx}{\textwidth}{@{}>{\raggedright\arraybackslash}p{3cm}>{\raggedright\arraybackslash}X>{\centering\arraybackslash}p{2.1cm}>{\centering\arraybackslash}p{2.1cm}>{\centering\arraybackslash}p{1.55cm}>{\centering\arraybackslash}p{1.8cm}>{\centering\arraybackslash}p{1.85cm}@{}}
\toprule
\multirow{3}{*}{\textbf{Primary design focus}}
& \multirow{3}{*}{\textbf{System}}
& \multirow{3}{*}{\shortstack{\textbf{Source--memory}\\\textbf{integration}}}
& \multirow{3}{*}{\shortstack{\textbf{Memory}\\\textbf{governance}}}
& \multicolumn{3}{c}{\textbf{Authorization continuity}} \\
\cmidrule(lr){5-7}
&
&
&
& \shortstack{\textbf{Transitive}\\\textbf{lineage}}
& \shortstack{\textbf{Policy}\\\textbf{composition}}
& \shortstack{\textbf{Policy}\\\textbf{re-evaluation}} \\
\midrule
 
\multirow{3}{=}{Personal memory}
& Letta, Mem0~\cite{memgpt,mem0}
& -- & -- & -- & -- & -- \\
& A-Mem, MIRIX~\cite{amem,mirix}
& -- & -- & -- & -- & -- \\
& AgentCore Memory, Memory Bank~\cite{agentcore_memory,vertex_memory_bank}
& -- & Partial & -- & -- & -- \\
 
\midrule
\multirow{2}{=}{Source-integrated memory}
& Zep~\cite{zep}
& Partial & -- & Core & -- & -- \\
& Oracle Agent Memory~\cite{oracle}
& Core & Partial & -- & -- & -- \\
 
\midrule
\multirow{2}{=}{Shared memory}
& G-Memory~\cite{g-memory}
& -- & -- & Partial & -- & -- \\
& SAP Organizational Memory~\cite{sap}
& -- & Partial & -- & -- & -- \\ 
 
\midrule
\multirow{3}{=}{Governed shared memory}
& MemClaw~\cite{caura_ai}
& -- & Core & Core & -- & Partial \\
& Collaborative Memory~\cite{collaborative_memory}
& -- & Core & Partial & Partial & Core \\
& MAP-Graph~\cite{map_graph}
& -- & Core & Core & Core & Partial \\ 
 
\midrule
Governed enterprise memory
& AkasicMEM & Core & Core & Core & Core & Core \\
 
\bottomrule
\end{tabularx}
 
\vspace{2pt}
\begin{minipage}{\textwidth}
\footnotesize
\textbf{Core} indicates a primary design target of the cited work.
\textbf{Partial} indicates that related mechanisms are provided but the requirement is not addressed across the complete source and memory lifecycle.
-- indicates that the requirement is outside the stated design scope of the cited work.
\end{minipage}
\vspace{-0.3cm}
\end{table*}

Because retrieved memory is itself an input to new formation, memory can be derived from memory as well as from sources.
Reflections synthesized from earlier memories~\cite{generative_agents} and insights abstracted from multi-agent interaction traces~\cite{g-memory} are examples.
In long-running agents, derivation continues across summaries, plans, and cached state.
Sources and memory are therefore connected by continuous derivation relations, so a lifecycle operation on one object affects the objects derived from it.
This consequence is sharpest for forgetting, because deleting a memory record while leaving its derived artifacts intact is, as recent work observes, a retrieval edit rather than a deletion~\cite{always_on_agents, execution_state_unlearning}.
An enterprise agent memory system must therefore manage the entire lifecycle over this structure rather than storage alone, and we formalize the structure in Section~\ref{sec:authorization_continuity}.

Agent memory is related to, but distinct from, long-context processing, context management, and retrieval-augmented generation (RAG).
Long-context processing expands how much information a model can exploit within a single inference or task~\cite{longbench}.
Context management compresses the information that accumulates over many interactions to sustain long-horizon tasks~\cite{mem1}.
Conventional RAG is a read-oriented pipeline that augments the model's input with information retrieved from an external store~\cite{rag}.
Agent memory, in contrast, decides which information becomes a persistent memory object, reuses it across sessions and tasks, and updates or removes it as new information arrives~\cite{agent_memory}.
In practice, these mechanisms can be combined, with long-context and context-management techniques serving as working memory and agent memory supporting memory over the long-term lifecycle.

\subsection{Existing Agent Memory Systems }
\label{sec:existing_memory_systems}

Existing agent memory systems differ in what their design primarily targets.
Memory formed by one principal may be reused only within that principal's own sessions, may be combined with the enterprise sources, or may be reused by multiple agents and users.
These design targets can overlap rather than define mutually exclusive capabilities.
Table~\ref{tab:existing_memory_systems} groups representative systems by their primary design focus and compares them against the three requirements introduced in Section~\ref{sec:introduction}, which are further developed in Section~\ref{sec:governed_enterprise_memory}.
The ratings reflect the design scope of the cited work rather than every feature of a product.

\vspace{-0.1cm}
\paragraph{Personal memory.}
These systems focus on memory for one principal and optimize continuity across that principal's sessions.
Letta moves information between a bounded context window and external storage~\cite{memgpt}, and Mem0 extracts salient facts from conversations and consolidates them for later retrieval~\cite{mem0}.
A-Mem and MIRIX organize memory into linked notes and typed components to improve recall~\cite{amem, mirix}.
Managed services such as Amazon Bedrock AgentCore Memory and Google's Memory Bank host this pattern for enterprises, scoping each memory to an identity or namespace and applying cloud access control at that scope~\cite{agentcore_memory, vertex_memory_bank}.
Isolation of this kind keeps one principal's memory apart from another's, but the policy on a memory is assigned by the developer rather than inherited from the data it was formed from.
Neither integration with enterprise sources nor governance of reuse across principals is within scope.

\vspace{-0.1cm}
\paragraph{Source-integrated memory.}
These systems cross the data boundary by combining agent memory with existing business data.
Zep ingests conversations and business data into a temporal knowledge graph whose facts link back to their episodes~\cite{zep}, but source-derived authorization does not follow that provenance: access to a memory is not composed from its sources and does not change when theirs does.
Oracle Agent Memory stores messages, profiles, and long-term memory in the same database as relational, JSON, and vector enterprise data and offers scope-aware retrieval~\cite{oracle}.
Authorization is applied per record by the database and application layers, but policy composition from contributing sources and exeplicit memory-to-memory derivation tracking are not specified.

\vspace{-0.1cm}
\paragraph{Shared memory.}
These systems cross the principal boundary without explicitly governing it.
G-Memory organizes multi-agent collaboration into interaction, query, and insight graphs so that lessons from past collaborations inform later tasks~\cite{g-memory}.
SAP's organizational memory captures knowledge produced by coding agents, promotes it to a curated store through a human approval gate, and serves it to subsequent agents~\cite{sap}.
Curation of this kind governs what enters memory, but neither system considers who may read a memory once it is shared or how the sources behind it constrain that reuse.

\vspace{-0.1cm}
\paragraph{Governed shared memory.}
These systems add policy to reuse across principals.
MemClaw provides scoped retrieval, temporal supersession, provenance tracking, and policy-governed propagation for fleets of agents~\cite{caura_ai}.
Collaborative Memory records the users, agents, and resources that contributed to each memory fragment and evaluates their current access conditions at retrieval time~\cite{collaborative_memory}.
MAP-Graph represents agents, sources, memories, and actions in a typed execution graph, traces the ancestry of each memory, and excludes records whose ancestry the requesting agent may not access~\cite{map_graph}.
Together these systems establish that provenance must be first-class and access evaluated under current conditions.
On those points, they address several of the requirements in Table~\ref{tab:existing_memory_systems}.
What remains outside their scope is source--memory integraion through a common data management and retrieval architecture that brings enterprise sources and agent memory together beyond source references and permission metadata.
They also do not fully support authorization continuity, with gaps in preserving derivation dependencies, composing inherited restrictions, or re-evaluating ancestor policies as permissions change.

Taken together, existing systems have developed the mechanisms needed for source--memory integration, memory governance, and authorization continuity within separate design scopes.
Governed shared memory systems come closest to the governance requirements.
No existing system, however, keeps enterprise sources and memory under one management boundary so that policy can be composed from sources through repeated derivation and re-evaluated as sources, principals, and policies change.
Section~\ref{sec:governed_enterprise_memory} develops the requirements of a system designed for this combined scope.

%% file: 3_section.tex
\vspace{-0.2cm}
\section{Governed Enterprise Memory}
\label{sec:governed_enterprise_memory}

Supporting source--memory integration alongside memory governance requires source authorization to remain effective after information is transformed into memory and reused.
Governed Enterprise Memory, introduced in Section~\ref{sec:introduction}, is agent memory designed around this joint lifecycle.
This paper focuses on authorization continuity across repeated memory derivation and changes in policy and requesting principals, and Section~\ref{sec:broader_governance} discusses the broader governance scope of semantic validity, retention and deletion, and auditability.
This section first motivates the authorization requirements of source-derived memory and then defines authorization continuity and its requirements.

\subsection{Authorization of Source-Derived Memory}
\label{sec:why_governed_enterprise_memory}

Existing approaches commonly control the reuse of memory through memory scopes such as a user, agent, thread, or tenant~\cite{oracle, caura_ai}.
Source--memory integration, however, raises a separate question, which is whether memory derived from enterprise sources can be governed by that scope alone.

\begin{figure}[t]
  \centering
  \includegraphics[width=0.95\linewidth]{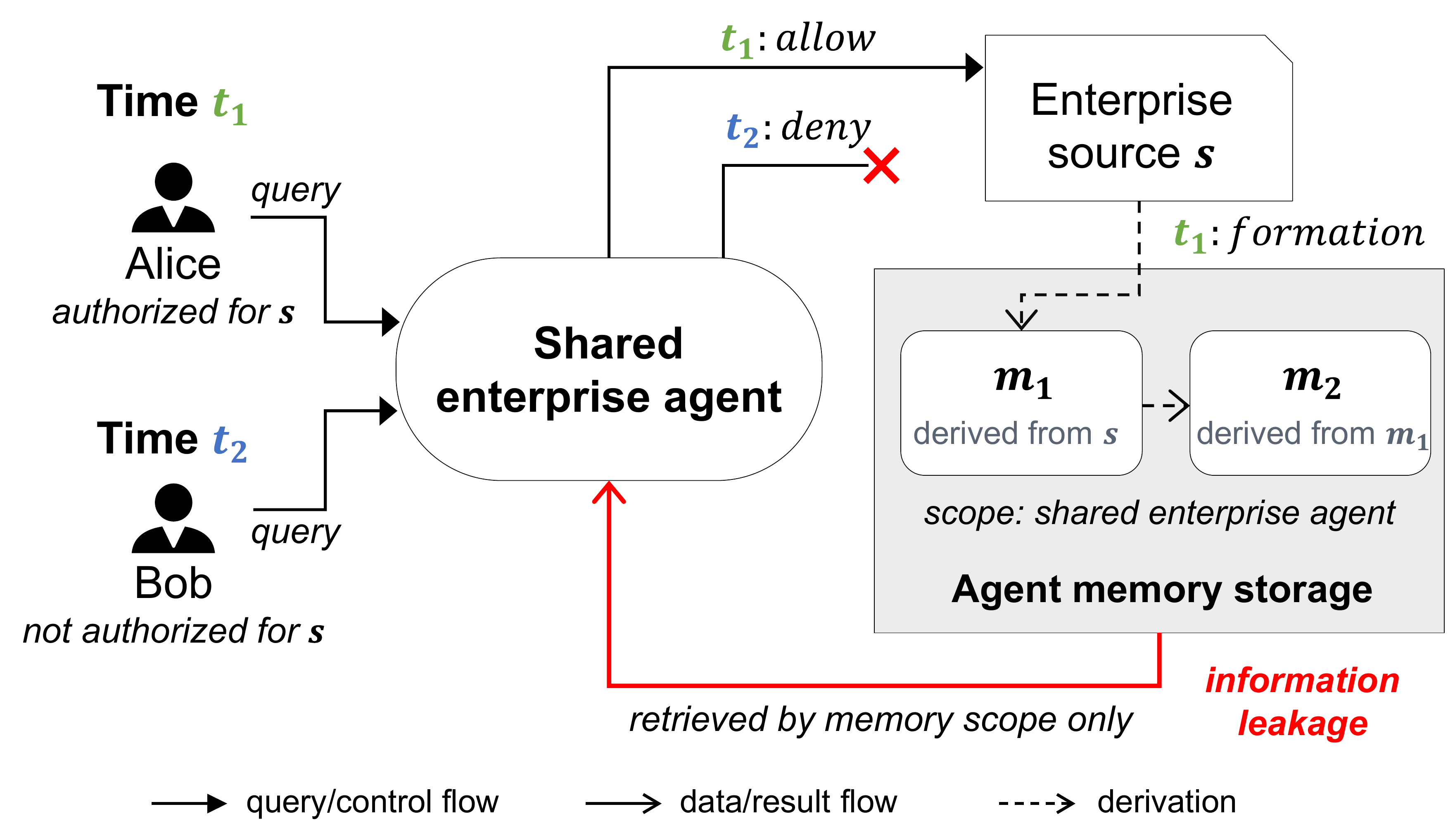}
  \vspace{-6pt}
  \caption{Example of source-to-memory authorization gap.}
  \label{fig:example_leakage}
  \vspace{-20pt}
\end{figure}

Figure~\ref{fig:example_leakage} illustrates this authorization gap with a simplified example.
At time $t_1$, a shared enterprise agent acting with Alice's authorization accesses enterprise source $s$ and forms memory $m_1$, from which it subsequently derives $m_2$.
Suppose that both memories are assigned the scope of the shared agent and can therefore be reused across its user sessions.
At time $t_2$ ($t_1 < t_2$), Bob invokes the same agent, which places his request within the scope of $m_1$ and $m_2$.
Because the agent now acts with Bob's authorization, however, the source system denies access to $s$.
If the memories are retrieved on the basis of their shared-agent scope alone, Bob obtains information from $s$ through $m_1$ or $m_2$ although he cannot access the source directly.
The source restriction has been bypassed through memory~\cite{collaborative_memory, map_graph}.

The example shows that an independently assigned memory scope is insufficient when it does not reflect the authorization policy of the source from which the memory was derived.
When enterprise sources and memory are managed together, source authorization must remain reflected as memory is formed, further derived, and retrieved.
The next subsection defines this requirement as \textit{authorization continuity}.

\subsection{Authorization Continuity}
\label{sec:authorization_continuity}

To define \textit{authorization continuity}, we first represent the lineage preserved across source-to-memory and memory-to-memory derivation.
Let $\mathcal{S}^t$ and $\mathcal{M}^t$ denote the sets of enterprise sources and memory objects registered up to time $t$.
Both are cumulative: an object enters when it is registered or formed and never leaves.
Forgetting can remove an object's content, embeddings, and eligibility for retrieval, but retains its identity, so it remains a vertex of the lineage graph.
The set of direct lineage edges recorded up to time $t$ is

\begin{equation}
\mathcal{D}^t \subseteq
\left(\mathcal{S}^t \cup \mathcal{M}^t\right)
\times \mathcal{M}^t,
\qquad
\mathcal{D}^{t} \subseteq \mathcal{D}^{t'} \;\text{ for } t \le t'.
\label{eq:derivation_edge}
\end{equation}

Each $(x,m)\in\mathcal{D}^t$ indicates that $x$, a source or an existing memory, directly contributed to the formation of $m$.
The relation captures source-to-memory lineage, memory-to-memory lineage, and formation from multiple inputs:
\[
s_1\rightarrow m_1,
\qquad
m_1\rightarrow m_2,
\qquad
\{s_1,s_2,m_1\}\rightarrow m_2,
\]
where the last is shorthand for the edges $(s_1,m_2)$, $(s_2,m_2)$, and $(m_1,m_2)$.
We also denote by $\mathrm{anc}^t(m)$ the sources and memories reachable from $m$ by traversing $\mathcal{D}^t$ backward.

We define \textit{authorization continuity} as the property that restrictions inherited from sources and parent memories continue to apply access to derived memory through repeated derivation and changes in requesting principals and managed policy state.
Maintaining this property imposes three requirements.

\vspace{-0.1cm}
\paragraph{Transitive lineage.}
The edges in $\mathcal{D}^t$ must be recorded at formation and preserved for the lifetime of every memory that depends on them, including after a contributing source or memory has been erased, so that $\mathrm{anc}^t(m)$ remains computable through any depth of derivation.
This preserves the provenance of derived memory and provides an auditable account of how each memory was formed.

\vspace{-0.1cm}
\paragraph{Policy composition.}
A new memory's policy must reflect both its intended scope, which specifies where it may be reused, and the restrictions of the sources and memories in its lineage, which constrain access to the information it inherits.
When several contribute, their restrictions must be combined, never replaced by an arbitrary policy or by the forming principal's own authorization.

\vspace{-0.1cm}
\paragraph{Policy re-evaluation.}
At retrieval, the policy composed at formation must be re-evaluated against the current principal and managed policy state, since the policies in its lineage may have changed.
The formation-time decision cannot simply be reused.
Re-evaluation must reflect both narrowing, when an authorization is revoked, and widening, when one is granted.

Together, these requirements motivate the following model of effective authorization.
Each object $x$ contributes one restriction of its own, $R^t(x)$: for a source, the set of principals its originating system currently authorizes; for a memory, the set of principals in the scope assigned to it at formation.
A deleted or deregistered source authorizes no one, so memory derived from it is withheld until explicitly declassified.
Both kinds of restriction are evaluated against the authorization graph as it stands at time $t$, so organizational changes take effect without any policy being rewritten.
Absent declassification, the \textit{effective authorization} of $m$ at time $t$ intersects its own restriction with those of everything it descends from:
\begin{equation}
\mathcal{A}^t(m) \;=\; R^t(m)\;\cap\;\bigcap_{x \,\in\, \mathrm{anc}^t(m)} R^t(x).
\label{eq:effective_authorization}
\end{equation}
In this equation, the intersection over $\mathrm{anc}^t(m)$ means the required access condition without requiring a full traversal of the lineage graph at every retrieval.
Declassification (Section~\ref{sec:governed_memory_formation}) is the one sanctioned exception: it removes a specific ancestor's restriction from the intersection for $m$, and is recorded rather than applied by editing $\mathcal{A}^t(m)$.

%% file: 4_AkasicMEM.tex
\section{AkasicMEM}
\label{sec:akasicmem}

\subsection{System Overview}
\label{sec:akasicmem_overview}

\begin{figure}[h]
  \centering
  \includegraphics[width=0.95\linewidth]{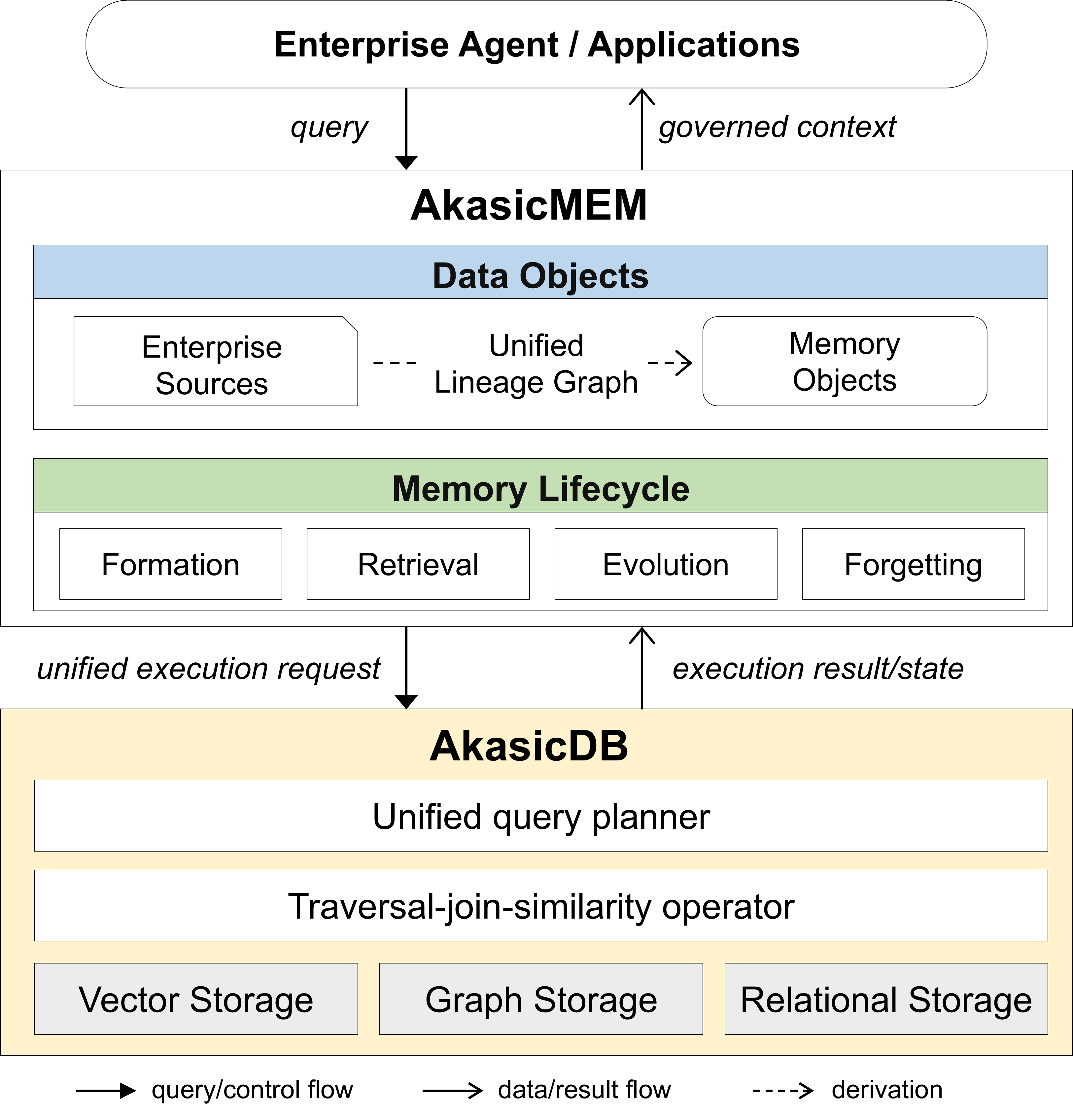}
  \vspace{-6pt}
  \caption{System overview of AkasicMEM on AkasicDB.}
  \label{fig:akasicmem}
  \vspace{-6pt}
\end{figure}

\textit{AkasicMEM} is a governed memory layer for enterprise agents and applications that realizes the authorization continuity defined in Section~\ref{sec:authorization_continuity}.
Figure~\ref{fig:akasicmem} shows its two parts.
The memory layer defines the governed data objects and the operations through which agents and applications form, retrieve, and manage memory.
Every operation runs on behalf of a requesting principal, which is a user or an agent acting with a user's authorization.
\textit{AkasicDB}, GraphAI's unified relational--graph--vector DBMS, holds the state of these objects and executes the operations~\cite{akasicdb}.

The memory layer manages three kinds of governed data objects.
An enterprise source is represented by a governed source object in AkasicDB: a stable identity linking it to its originating system, which remains the system of record, the content needed for joint retrieval, and its registered authorization state.
Memory objects are the units of memory formed during task execution, each with a persistent identity, content and corresponding vector representations, an assigned memory scope, and a composed authorization policy computed during formation.
The lineage graph records the direct derivation relation $\mathcal{D}^t$ of Section~\ref{sec:authorization_continuity} between sources and memory objects.
AkasicMEM can also maintain further data objects for governance such as the authorization graph and decision records.
The authorization graph records the principals with hierarchical organizational units and authorization scopes that policies define, along with the relationships that determine which principal holds which scope.
Decision records capture each declassification and each retrieval decision with its basis.
Sources and memory objects thus keep distinct identities and policies, while the lineage graph connects them and the authorization graph provides the organizational relationships required to evaluate those policies.

\begin{figure}[t]
  \centering
  \includegraphics[width=0.99\linewidth]{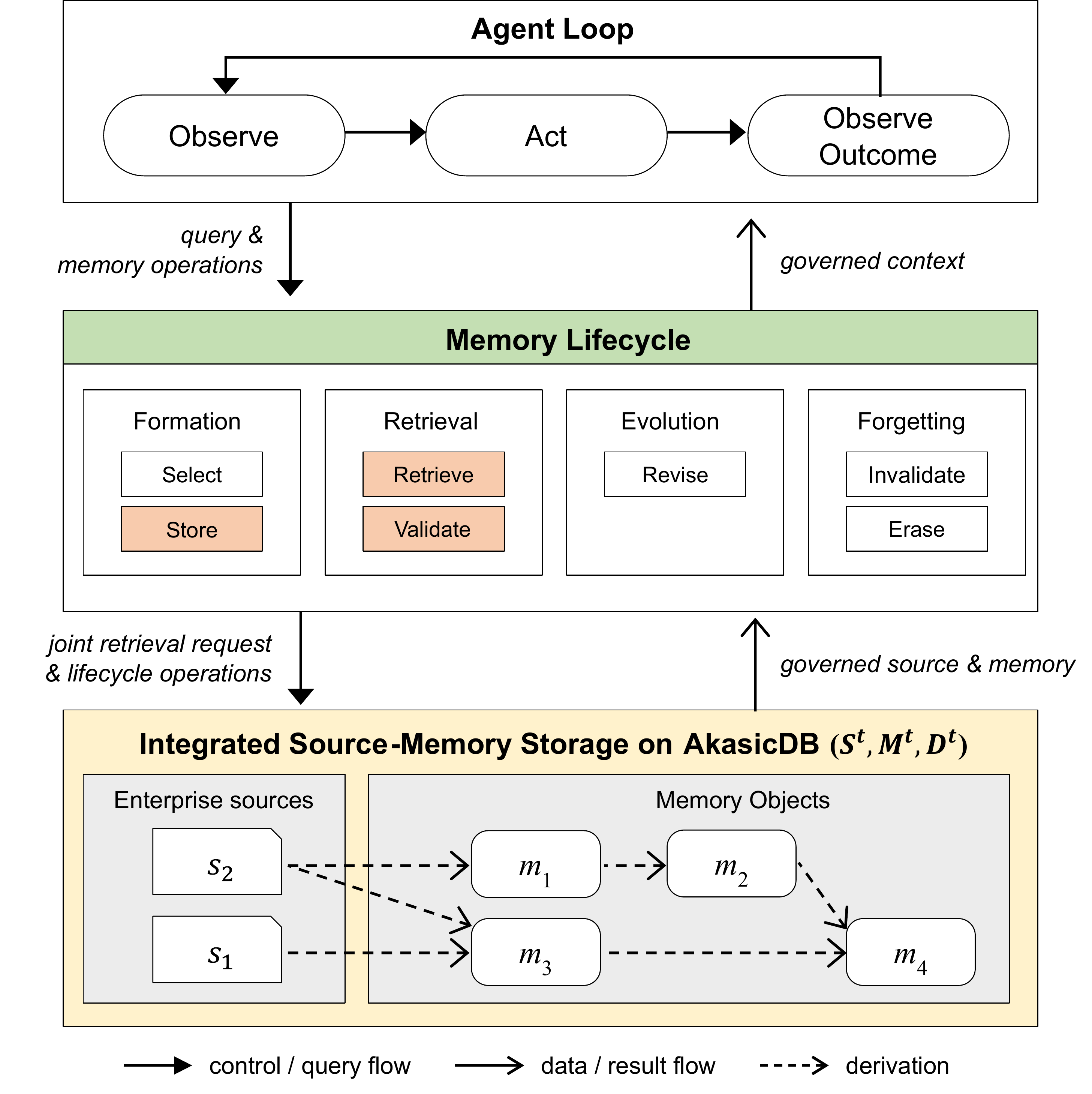}
\vspace{-6pt}

  \caption{Unified source--memory storage and memory lifecycle management in AkasicMEM.}
  \label{fig:lifecycle}
  \vspace{-12pt}
\end{figure}

Figure~\ref{fig:lifecycle} shows how AkasicMEM coordinates the lifecycle dynamics of Section~\ref{sec:agent_memory_lifecycle} with joint retrieval over integrated source--memory storage.
At the top, the agent loop of observing, acting, and observing the outcome issues memory operations and receives governed context, which can feed the next formation.
In the middle, the memory layer exposes the four dynamics as first-class operations.
Formation consists of \textit{Select}, which determines what information is worth retaining, and \textit{Store}, which materializes it as a memory object.
Retrieval consists of \textit{Retrieve}, which identifies task-relevant candidates, and \textit{Validate}, which determines whether each candidate may be used in the current context. Evolution modifies existing memory through \textit{Revise}, and forgetting either excludes memory from further use through \textit{Invalidate} or removes its content from storage through \textit{Erase}, retaining its identity in the lineage graph as Section~\ref{sec:authorization_continuity} requires.
At the bottom, an \textit{integrated source--memory storage} holds the enterprise sources $\mathcal{S}^t$, the memory objects $\mathcal{M}^t$, and the lineage edges $\mathcal{D}^t$ among them in one store, so that memory derived from a source, or from other memory, remains connected to its origins, as $m_4$ is to $s_1$ and $s_2$ through $m_3$ in the figure.
Lifecycle operations read from and write to this shared storage, the concrete form of source--memory integration that takes place in AkasicMEM.


This paper focuses on \textit{Store}, \textit{Retrieve}, and \textit{Validate}, the operations through which authorization passes from sources into memory and from memory to a requesting principal.
\textit{Store} records a lineage edge from every contributing source and memory and composes the new memory's policy from its scope and the restrictions of those inputs (Section~\ref{sec:governed_memory_formation}).
\textit{Retrieve} and \textit{Validate} execute as one governed query that selects semantically relevant candidates and evaluates each against the current policy state and principal, so only authorized memory enters the agent's context (Section~\ref{sec:governed_memory_retrieval}).
A change to a source or memory policy travels through the lineage graph as a flag on downstream memories, which \textit{Validate} resolves when the affected memory is next selected as a candidate.
\textit{Select} and \textit{Revise} determine what is retained and how it evolves, \textit{Invalidate} and \textit{Erase} its later disposition; Section~\ref{sec:discussion} discusses them.
AkasicDB executes these operations as one workload, holding sources, memory metadata, policies, and decision records as relational data, lineage and authorization as graph data, and content embeddings as vector indexes, under a single query planner and transaction manager (Section~\ref{sec:unified_execution}).


\subsection{Governed Memory Formation}
\label{sec:governed_memory_formation}

Within memory formation, AkasicMEM governs the \textit{Store} operation, which realizes the first two requirements of authorization continuity, transitive lineage and policy composition.
For each newly formed memory, AkasicMEM records how it was derived and determines its authorization policy from both its intended memory scope and the restrictions of its inputs.

For transitive lineage, AkasicMEM records a direct lineage edge from every contributing enterprise source and existing memory to the newly formed memory.
It maintains these edges in a \textit{lineage graph} whose vertices represent enterprise sources and memory objects and whose edges represent the direct lineage relation $\mathcal{D}^t$.
Edges form source-to-memory and memory-to-memory paths, so every memory remains traceable to the enterprise sources that contributed to it.
The lineage graph thereby preserves the transitive derivation relationships that subsequent policy evaluation and audit require.

For policy composition, \textit{Store} assigns the new memory its scope and computes its effective authorization by Eq.~\eqref{eq:effective_authorization} at formation time.
It follows the lineage graph to identify the contributing inputs and their restrictions, and uses the authorization graph to interpret the organizational scopes that these restrictions name.
The result is stored with the memory as its \textit{composed policy}; this stored value is a cache of Eq.~\eqref{eq:effective_authorization}, and Section~\ref{sec:governed_memory_retrieval} describes when it is recomputed.
Because the composed policy intersects every restriction imposed by the contributing inputs rather than selecting any single one, it over-restricts a derived memory whose content no longer retains restricted information from an input.
AkasicMEM relaxes such a memory through explicit \textit{declassification} with respect to a specific ancestor, typically a source, which records that ancestor as excluded from the intersection in Eq.~\eqref{eq:effective_authorization} for the memory.
Memory later derived from it inherits the exclusion unless it depends on that ancestor through another lineage path.
Declassification proceeds either automatically, through semantic verification, or manually, with the approval of an authorized human reviewer.
The automated path uses natural language inference (NLI)~\cite{nli} to estimate whether restricted information from the ancestor remains in the derived memory.
On either path, the lineage edge itself is retained and the decision and its basis are recorded for audit.
Because a declassification is a recorded exclusion rather than an edit to the composed policy, it is preserved by later recomposition and is not undone by a subsequent change to the ancestor's policy.

\subsection{Governed Memory Retrieval}
\label{sec:governed_memory_retrieval}

Memory retrieval in AkasicMEM realizes the third requirement of authorization continuity, policy re-evaluation.
Governed formation stores a composed policy for each memory, and governed retrieval determines whether the current requesting principal belongs to the memory's effective authorization $\mathcal{A}^t(m)$ under the current policy state.
AkasicMEM performs this evaluation as part of retrieval and returns only the memories that their current policies permit.

A straightforward implementation would propagate every policy change through the lineage graph and immediately recompute the composed policies of all affected memories.
The cost of such eager propagation grows with the number of downstream memories, most of which may never be retrieved. AkasicMEM therefore adopts lazy \textit{retrieval-time policy revalidation}.
When the policy of an enterprise source or memory changes, AkasicMEM follows the lineage graph and only flags the potentially affected downstream memories, marking their composed policies as stale, without eager recomputation.

When a flagged memory becomes a retrieval candidate, AkasicMEM recomputes $\mathcal{A}^t(m)$ by Eq.~\eqref{eq:effective_authorization}, following lineage to the current restrictions of its contributing sources and ancestor memories and honoring recorded declassifications, and replaces the stale composed policy with the result.
The candidate is then served to the requesting principal $p$ only if $p \in \mathcal{A}^t(m)$.
Because the recomputation uses the current policy state rather than the fixed state recorded at formation, a change of authorization can narrow or widen access.

To reduce the overhead of lineage traversal, AkasicMEM periodically revalidates flagged candidates in batches rather than one memory at a time. Such batched policy recomposition can be optimized by grouping and parallelization.
With such retrieval-time revalidation, only memory candidates that remain authorized are returned, while policies marked as stale are revalidated before use.


\subsection{Unified Execution on AkasicDB}
\label{sec:unified_execution}

The governed formation and retrieval of Sections~\ref{sec:governed_memory_formation} and \ref{sec:governed_memory_retrieval} form a mixed relational, graph, and vector workload.
The \textit{Store} operation creates a memory object and its vector representation while recording lineage edges and composing its policy from the restrictions of its inputs.
Retrieval combines semantic candidate selection with metadata filtering, lineage traversal, and evaluation under current policy.
If relational, graph, and vector data were managed in separate systems, these operations would have to be decomposed into independently executed queries.
The application would then transfer intermediate results between systems, coordinate their execution, and maintain consistency across separate transaction boundaries.
Without additional synchronization, updates to policy and lineage state could leave memory and its applicable authorization in inconsistent views.

AkasicMEM instead executes this workload on GraphAI's AkasicDB~\cite{akasicdb}, a unified data substrate.
AkasicDB builds on a native graph--relational processing architecture~\cite{chimera} and extends it with vector data and similarity search.
It manages relational tables, graph topology, and vector indexes under a common transaction and combines relational filtering, graph traversal, and vector search within a single query plan.
A unified planner can therefore optimize the whole workload rather than coordinate independently optimized subqueries, which avoids application-level round trips and the materialization of intermediate candidate, lineage, and policy results.

During governed formation of a memory object, its vector representation, lineage edges, composed policy, and any declassification record are written in one transaction.
The new memory therefore becomes visible only together with the lineage and the composed policy state.
The same holds for a policy change within AkasicDB: subsequent retrieval observes the updated state and flags without cross-store delay.
For a source whose system of record is external, authorization changes must be synchronized into the registered policy state; that synchronization is an integration requirement whose latency bounds how promptly Eq.~\eqref{eq:effective_authorization} reflects the change.

During governed retrieval, AkasicDB combines vector-based candidate selection, relational filtering, lineage traversal, and authorization evaluation in a single execution plan.
The candidate set produced by semantic search is constrained directly by structured conditions and graph relationships, without any data transfer to a separate system. Also, batch revalidation of flagged candidates runs as a set operation over that candidate set rather than as a request per memory.
Because authorization is enforced within the database, candidates that fail the current policy are filtered out before any content reaches the agent or application layer.

Unified execution does not remove the cost of semantic search, lineage traversal, or policy evaluation.
It allows these operations to be optimized and executed jointly, without costly and tedious cross-system coordination.
Single-plan query processing and a common transaction boundary are the execution foundation on which AkasicMEM applies authorization continuity without separating retrieval from lineage and policy processing.

%% file: 5_discussion.tex
\vspace{-0.2cm}
\section{Discussion and Future Work}
\label{sec:discussion}

Authorization continuity addresses one way in which shared memory fails, the leakage of information to principals who may not access its sources.
Work on governed shared memory identifies three further failure modes, namely memory read after an update has made it stale, conflicting memories left active together, and memory that can no longer be traced to its origin~\cite{caura_ai}.
The lineage that AkasicMEM maintains to carry policy from sources to derived memory (Section~\ref{sec:akasicmem}) is also the structure through which these must be handled.
We discuss how governance extends along that lineage, how the design can be extended toward useful as well as governed memory, and how such a system should be evaluated.

\subsection{\scalebox{0.95}{Governance Beyond Authorization Continuity}}
\label{sec:broader_governance}

\paragraph{Validity under source change.}
When a source is corrected or retracted, the memory derived from it remains authorized for the same principals yet may no longer be true, and no access decision can detect this.
Lineage should therefore carry validity as well as policy, so that a change to a source marks its descendants for revision or invalidation, much as a policy change flags them for re-evaluation.
The two differ in that a change of scope composes mechanically, whereas a change of content requires judgment, because a summary may be unaffected by the correction of one fact in its sources.
Propagation can flag or suspend derived memory, but resolving it requires re-verification against evidence or a person.
Revisions should be recorded as new versions rather than as overwrites, which preserves what was believed and served in the interim.
The same mechanism should apply when a contributor is later found unreliable, so that lineage carries trust in an origin alongside authority over it.

\vspace{-0.1cm}
\paragraph{Retention and deletion through lineage.}
Retention and deletion obligations attach to sources but must reach everything derived from them, and lineage makes this transitive part tractable.
Three difficulties remain.
Derivation is many-to-one, so a memory formed from several sources need not be deleted when one is.
Deciding whether it still depends on the deleted content is the same judgment that declassification makes (Section~\ref{sec:toward_usefulness}).
Deletion must be verifiable, which in a system with vector retrieval means probing for memory semantically close to the deleted content rather than checking that a record is gone~\cite{execution_state_unlearning, always_on_agents}.
Sources carry different retention periods, and derived memory should inherit the most restrictive by the same composition that governs its authorization.
Underlying all three is a tension with usefulness, because aggressive deletion destroys the knowledge that motivates enterprise memory.
A lesson whose value has outlived its source should be declassified deliberately rather than exempted from deletion.

\vspace{-0.1cm}
\paragraph{Auditability.}
For each decision, an organization should be able to establish who asked, what was served or withheld, and under which policy, without disclosing content to the auditor.
The current design records each decision and its basis, but two requirements go beyond logging.
Reproducing a past decision requires the authorization graph and source policies as they stood at the time, so policy must be versioned alongside memory.
Versioned policy also allows a proposed change to be simulated against current lineage before it is applied.
When a source is revoked, memory already served to newly unauthorized principals cannot be retracted, but lineage joined with the recorded decisions can report to whom and when such disclosure occurred.
The auditor's own access is itself an authorization, because audit records reveal the existence and reach of memory even without its content.

\vspace{-0.2cm}
\subsection{Toward Useful Governed Memory}
\label{sec:toward_usefulness}

Enterprise agent memory can be characterized by a progression of requirements, from formation and retrieval, through provenance and enforcement, to controlled declassification and adaptation from outcomes.
This paper concentrates on provenance and enforcement, and its formation and retrieval are deliberately conservative.
Governance becomes binding when every employee works with a paired agent, because memory must then inherit the person's own authorizations.
At that scale, what to remember and what to return also become the dominant questions.
Usefulness is nonetheless not a second axis beside governance, because selection, consolidation, and decay create or remove derivation and so run inside the same lineage that carries policy.

\vspace{-0.1cm}
\paragraph{Selection.}
\textit{Select} is outside this paper's scope, and admitting a memory because evidence supports it tests whether the memory is true but not whether it is likely to be useful again.
Whether the acting agent~\cite{mem1} or an observer outside the task~\cite{sap} should judge future usefulness is contested.
Governed memory favors the observer at formation, because the memory system must assign each memory's authorization from its lineage, and that assignment cannot be left to an agent that sees only its own task.
The practical pattern is to filter early, store observations close to their evidence, and consolidate offline, keeping online and consolidated stores distinct~\cite{lightmem}.
Consolidation exposes a tension particular to governed memory, because a memory consolidated from many sources absorbs the restrictions of all of them and becomes more restricted the more broadly useful it is.
Integrating consolidation into governed formation is ongoing work.

\vspace{-0.1cm}
\paragraph{Retrieval and forgetting.}
Governed retrieval returns the memory most similar to the query among those a principal may see, evaluating current policy in the same query.
It does not yet consider how useful a memory has proved in practice.
Recent work assigns each memory a retention score that increases when the memory is retrieved and used and decreases when it is retrieved but not used, and finds this preferable to decay based on age alone~\cite{wmt}.
A side effect is that injected content that is never used loses influence over time.
Such decay also supplies the usefulness-based trigger for invalidation that the current design lacks.
If invalidation operates over lineage rather than by deleting records, decayed memory can leave retrieval while its derivation remains for audit.
Governed retrieval could also exploit a signal that ungoverned systems lack, because a memory repeatedly relevant yet repeatedly filtered out marks knowledge blocked by governance rather than irrelevance, and is a natural candidate for declassification review.

\vspace{-0.1cm}
\paragraph{Controlled declassification.}
Authorization continuity keeps a derived memory at least as restricted as its sources.
Controlled declassification is the authorized exception, under which a derived memory receives a wider audience because its content no longer depends on the restricted material.
A procedural lesson learned in a confidential engagement is the typical case, because its value elsewhere is high and it discloses nothing of the source.
The current design gates declassification on a natural language inference test of whether restricted information from the ancestor remains in the derived memory, or on the approval of an authorized reviewer (Section~\ref{sec:governed_memory_formation}).
The automated test is necessary but not sufficient, because a text may be independent of a source sentence by sentence yet disclose it in aggregate.
We therefore expect the automated path to serve as a screen that informs human approval rather than as a substitute for it.

\vspace{-0.1cm}
\paragraph{Outcome feedback.}
The current design does not yet adapt from outcomes, which would close the loop from observed outcomes back to selection and retrieval so that utility and trust reflect experience rather than a single assignment.
A record that a memory was useful to a principal is itself derived information, and can disclose for instance that knowledge from one engagement was applied in another.
Feedback must therefore be recorded and consumed in the same content-free, scoped manner as audit records.

\subsection{Evaluating Governed Enterprise Memory}
\label{sec:evaluation}

Existing benchmarks measure recall of a single user's conversational history~\cite{locomo, longmemeval, memoryagentbench}, with no sources, one principal, and no change of policy, so a system that leaks across principals scores as one that does not.
Evaluations of governed shared memory add principals and permissions~\cite{collaborative_memory, map_graph}, but their resources live inside the memory layer, while derivation depth and policy change are fixed rather than varied.
An evaluation of governed enterprise memory must therefore be constructed.
It needs a synthetic organization with sources, principals, and authority relationships, a task stream that forms and derives memory, and a timeline of governance events such as revocations, corrections, and deletions.

Governance correctness is measured by leakage, the fraction of serves that the current policy of some source in the served memory's lineage excludes, reported against derivation depth and time since policy change.
Companion measures cover the other three failure modes, together with incomplete propagation after a governance event and unverifiable forgetting.
The cost to usefulness is measured by over-restriction, the fraction of serves withheld although current policy permits them, and by retrieval relevance, downstream utility, freshness, and trust calibration~\cite{sap}, each compared for authorized principals with governance enabled and disabled.
Overhead is measured at formation, retrieval, and propagation against an ungoverned baseline as the number of principals, the depth of the authorization graph, derivation depth, and memory volume grow.

Three difficulties complicate these measurements.
Leakage is semantic, because a paraphrase discloses its source as surely as a copy, so ground truth requires judged comparison or adversarial probing that attempts to reconstruct a restricted source from what a principal may retrieve~\cite{agentpoison}.
The same probing tests declassification, whose failure is disclosure in aggregate.
The measures trade against one another, because a system can eliminate leakage by withholding everything, so a system is characterized by its trade-off among leakage, over-restriction, and cost rather than by a single figure.
Comparing that trade-off against ungoverned, scope-isolated, and governed shared baselines shows what source integration contributes rather than assuming it.
Checking a decision against the policy in force when it was made requires the versioned policy state of Section~\ref{sec:broader_governance}, which makes reproducible evaluation and auditability the same requirement.

%% file: 6_conclusion.tex
\section{Conclusion}
\label{sec:conclusion}
 
Memory derived from enterprise sources carries their information beyond the authorization under which it was obtained, and a memory scope alone cannot tell whether a later requester may see it.
We defined authorization continuity as the property that source restrictions remain effective as memory is repeatedly derived and reused under changing policies and principals, and presented AkasicMEM, which realizes it by recording lineage at formation, composing each memory's policy from its sources, and re-evaluating that policy at retrieval time over an integrated source--memory storage on AkasicDB.
Across the directions discussed in Section~\ref{sec:discussion}, Governed Enterprise Memory emerges less as a set of features than as a single dependency structure managed across the lifecycle.
We expect the tension between accumulated restrictions and the value of shared knowledge, which controlled declassification is meant to release, to be the central design question for enterprise agent memory.